\documentclass[useAMS,usenatbib]{mn2e}

\usepackage{natbib}

\usepackage{latexsym,graphicx}
\usepackage{color}
\usepackage{fixltx2e}
\usepackage{verbatim} \usepackage{float}
\usepackage{amsmath,amssymb}
\usepackage{times}

\usepackage{setspace}
\usepackage{rotating}
\usepackage{pdflscape}
\usepackage{subfig}

\usepackage[squaren, Gray, cdot]{SIunits}

\def\apgt{\ {\raise-.5ex\hbox{$\buildrel>\over\sim$}}\ }
\def\aplt{\ {\raise-.5ex\hbox{$\buildrel<\over\sim$}}\ }

\let\oldhat\hat
\renewcommand{\hat}[1]{\oldhat{\mathbf{#1}}}

\title[Accretion-driven bursts in massive star formation]{
\textcolor{black}{On the existence of accretion-driven bursts in massive star formation}
}

\author[D. M.-A.~Meyer et al.]
       {D. M.-A.~Meyer,$^{1}$\thanks{E-mail: dominique.meyer@uni-tuebingen.de} E. I.~Vorobyov,$^{2,3}$ R.~Kuiper$^{1}$ and W.~Kley$^{1}$\\
       $^{1}$Institut f\" ur Astronomie und Astrophysik, Universit\" at T\" ubingen,  Auf der Morgenstelle 10, 72076 T\" ubingen, Germany\\
       $^{2}$Department of Astrophysics, The University of Vienna, Vienna, A-1180, Austria\\
       $^{3}$Research Institute of Physics, Southern Federal University, Stachki 194, Rostov-on-Don, 344090, Russia\\
       }

\begin{document}

\date{Received June 2016; accepted Month day Year}

\maketitle
   
\label{firstpage}

\begin{abstract}
Accretion-driven luminosity outbursts are a vivid manifestation of variable mass 
accretion onto protostars. They are known as the so-called FU Orionis phenomenon in the 
context of low-mass protostars. More recently, this process has been found in models 
of primordial star formation. Using numerical radiation hydrodynamics 
simulations, we stress that present-day forming massive stars also experience 
variable accretion and show that this process is 
accompanied by luminous outbursts induced by the episodic accretion of gaseous 
clumps falling from the circumstellar disk onto the protostar. 
Consequently, the process of 
accretion-induced luminous flares is also conceivable in the high-mass regime of 
star formation and we propose to regard this phenomenon as a general mechanism 
that can affect protostars regardless of their mass and/or the chemical properties 
of the parent environment in which they form. 
In addition to the commonness of 
accretion-driven outbursts in the star formation machinery, we conjecture that 
luminous flares from regions hosting forming high-mass star may be an observational 
implication of the fragmentation of their accretion disks. 
\end{abstract}



\section{Introduction}
\label{sect:introduction}

Stars form in reservoirs of gas and dust, which collapse under their own gravity.
However, a significant fraction of pre-stellar gas,
thanks to the conservation of the gas net angular momentum, 
lands onto a centrifugally balanced circumstellar disc rather than falling directly onto the 
forming protostar.  The manner in which matter accretes from the disc onto the 
star is still poorly understood and, notably, early spherical collapse 
models~\citep{larson_mnras_145_1969,shu_apj_214_1977} that neglect the disc 
formation phase cannot explain the variety of mass accretion rates observed in 
present-day star-forming regions~\citep{vorobyov_apj_704_2009}. In particular, 
these models yield accretion luminosities that are factors of 10$-$100 greater 
than the mean luminosity measured for nearby star-forming regions.

This so-called "luminosity problem"~\citep{kenyon_aj_99_1990} can be solved if the accretion history onto 
the protostar is not smooth or quasi-constant, as predicted by spherical 
collapse models, but highly time-variable, as it naturally occurs in 
self-consistent models that follow the transition from clouds to protostellar 
discs~\citep{dunham_apj_747_2012}. In these models, protostars spend most of 
their time in the quiescent phase with low rate of accretion, which is 
interspersed with short but intense accretion 
bursts~\citep[see][]{voroboyov_apj_650_2006,vorobyov_apj_714_2010, 
machida_apj_729_2011, zhu_apj_746_2012,vorobyov_apj_805_2015}. Spectacular 
examples of these burst systems are a special class of  young stars called FU 
Orionis objects, which display outbursts of a factor 
of hundreds in luminosity which last several decades to hundreds of years. Such 
flares are thought to be due to drastic increases in the mass accretion rate of 
such young stars~\citep{kley_apj_461_1996}.

Until recently, it was thought that FU-Orionis-type accretion and luminosity bursts were constrained 
to occur in the solar mass regime of present-day star formation~\citep{audard_2014}. However, recent numerical 
hydrodynamics simulations of primordial disc formation around the first very 
massive stars have also revealed the presence of accretion bursts caused by disc 
gravitational fragmentation followed by rapid migration of the fragments onto 
the protostar~\textcolor{black}{\citep{stacy_mnras_403_2010,greif_mnras_424_2012,smith_mnras_424_2012,
vorobyov_apj_768_2013,hosokawa_2015}}. These studies 
have revealed highly variable protostellar accretion with multiple bursts, 
exceeding in numbers their present-day 
counterparts~\citep{desouza_mnras_540_2015}. The same process of bursts driven 
by disc fragmentation operates around primordial super-massive stars, relaxing the 
ultraviolet photon output and enabling the stellar growth to the limit where 
general-relativistic instability results in the formation of super-massive black 
holes~\citep{sakurai_mnras_549_2016}.

\begin{figure*}
        \centering
        \begin{minipage}[b]{ 0.245\textwidth}
                \includegraphics[width=1.0\textwidth]{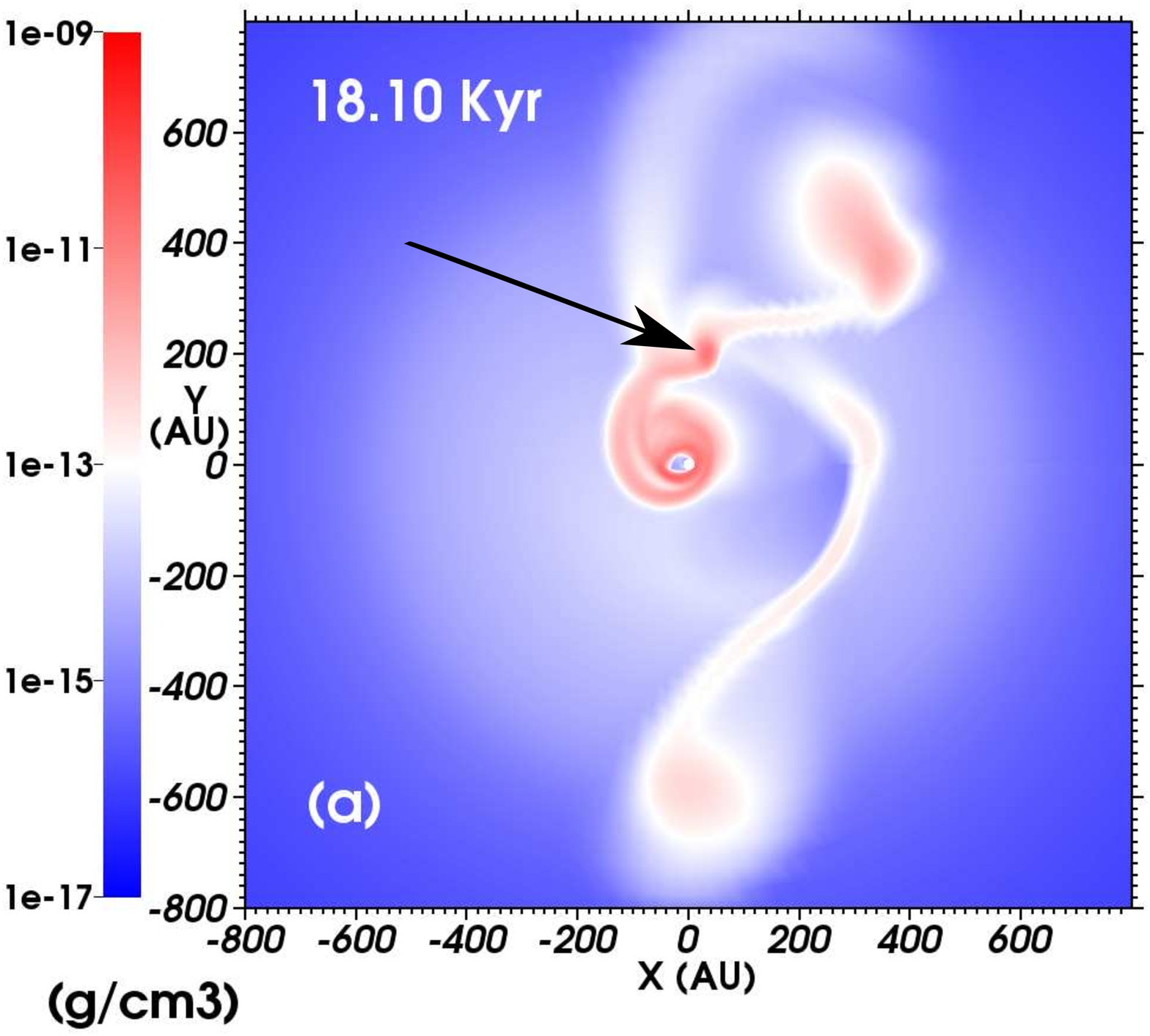}
        \end{minipage} 
        \begin{minipage}[b]{ 0.245\textwidth}
                \includegraphics[width=1.0\textwidth]{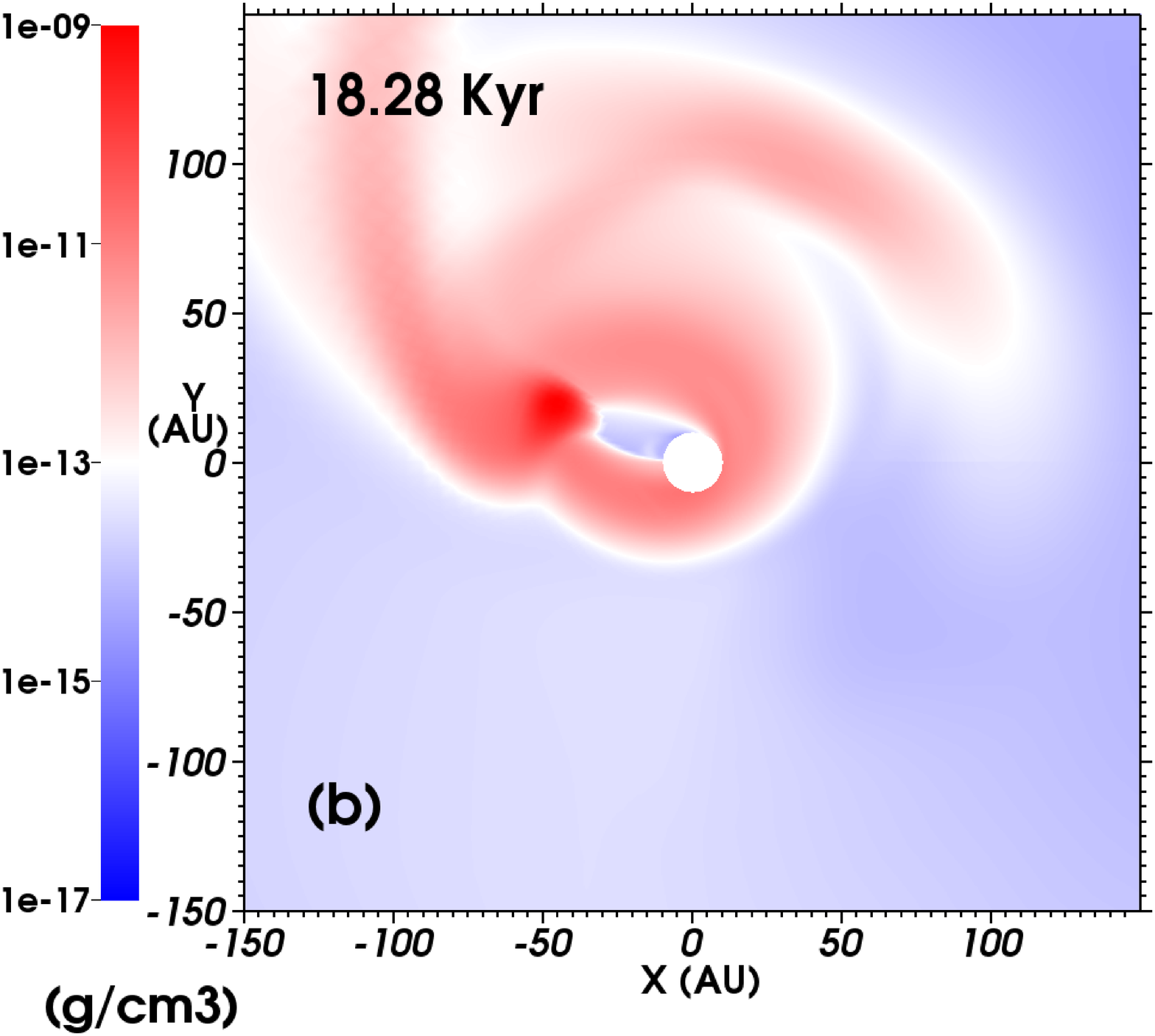}
        \end{minipage}  
        \begin{minipage}[b]{ 0.245\textwidth}
                \includegraphics[width=1.0\textwidth]{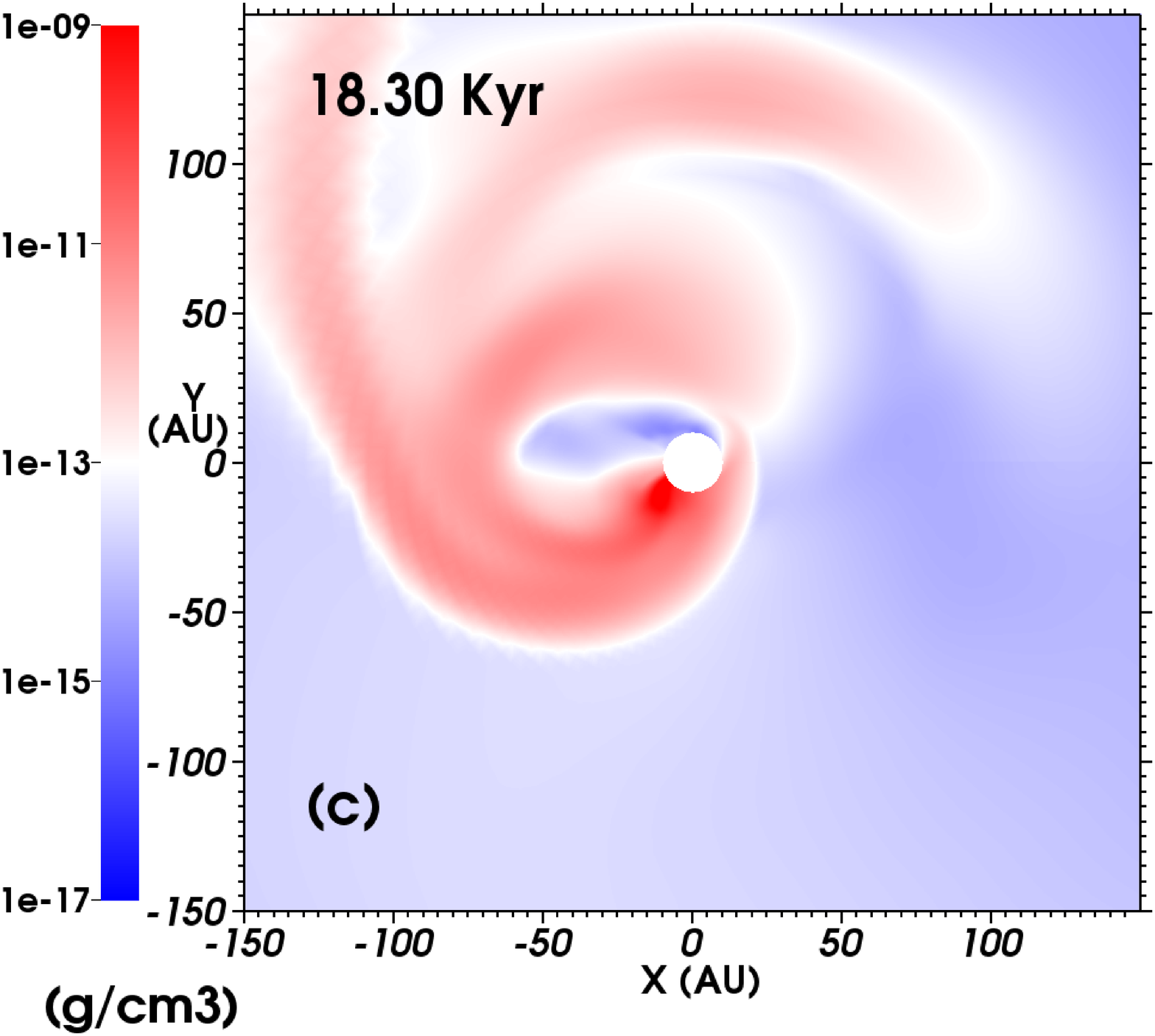}
        \end{minipage} 
        \begin{minipage}[b]{ 0.245\textwidth}
                \includegraphics[width=1.0\textwidth]{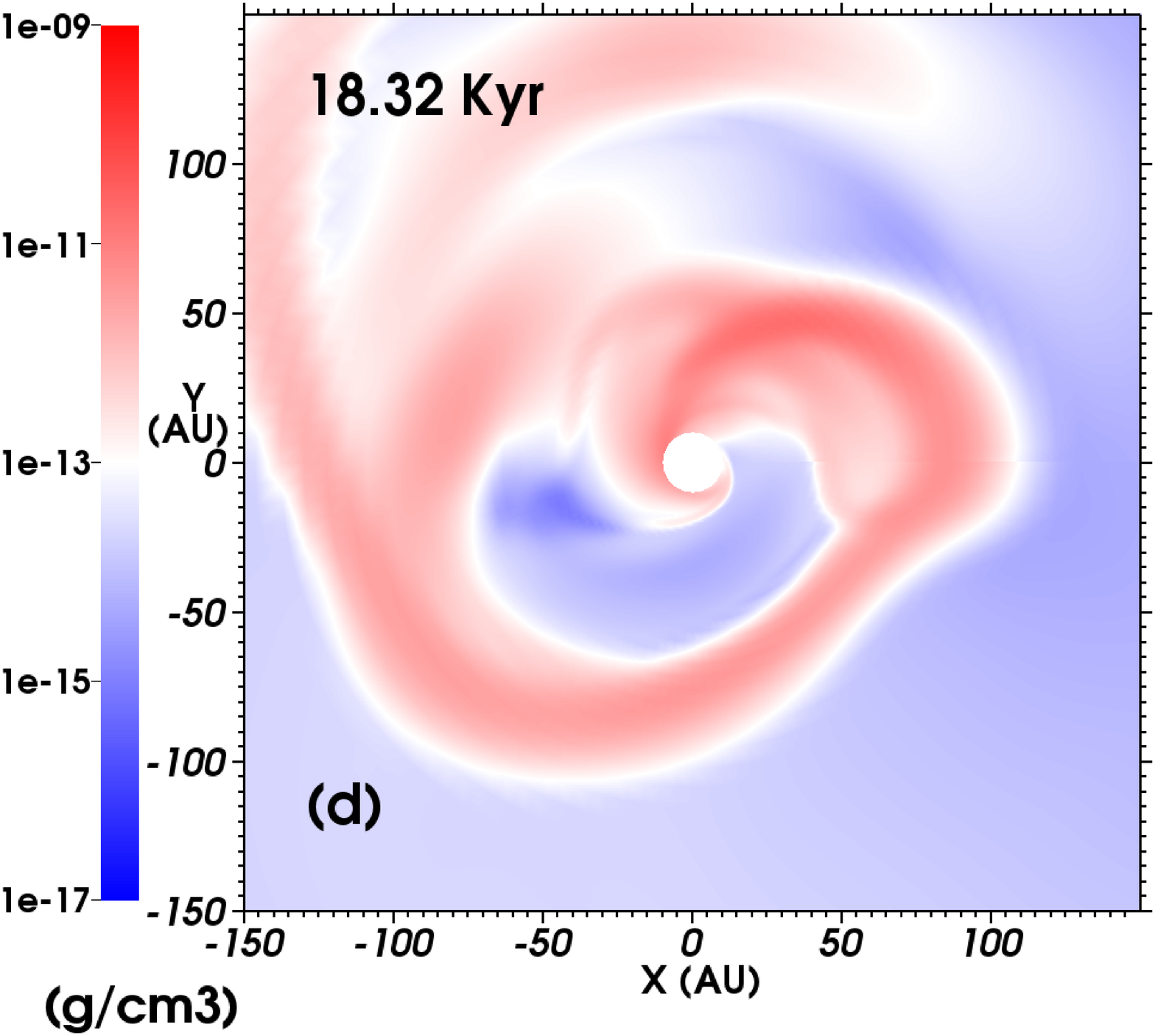}
        \end{minipage}
        \caption{ 
                 Midplane density \textcolor{black}{in the center of the computational domain around} the time of the
first outburst. (a) \textcolor{black}{The region within 800~AU}
when a clump forms in a spiral arm $\sim200\, \rm AU$ \textcolor{black}{away from} the protostar, 
at a time $18.10\, \rm kyr$. Panel (b-c) display zooms to illustrate the migration and accretion of a part of
the clump at times $18.28$, $18.30$ and $18.32\, \rm kyr$,
respectively. The density is plotted in g/cm$^{3}$ on a logarithmic
scale and the size of the panels is in AU.        
                 }      
        \label{fig:disk_density_plots}  
\end{figure*}

Consequently, the emerging question is, how universal is variable accretion with 
episodic bursts in star formation and whether it can be associated to a unique 
physical mechanism? Time-variability of accretion onto present-day massive 
stars at the early phase of their formation is a known 
\textcolor{black}{
process~\citep{krumholz_apj_656_2007,peters_apj_711_2010,kuiper_apj_732_2011,klassen_arXiv160307345K}. 
Those studies interpret this phenomenon as a natural consequence of the
three-dimensional nature of their self-gravitating numerical simulations,
while~\citet{kuiper_apj_772_2013} explains it by the interplay between mass 
accretion, stellar evolution, and radiative feedback. 
}
They report how asymmetries 
can develop in self-gravitating discs and generate an azimuthal anisotropy in 
the accretion flow onto the protostars. One can particularly notice that in 
addition to its variable character, it is interspersed with several accretion 
peaks~\citep[see fig.~4 of][]{klassen_arXiv160307345K}. 
In the above cited references, the sharp increases of the accretion rate are generated in 
simulations assuming different pre-stellar core masses, ratio of kinetic by 
gravitational energy $\beta$, assuming either a {\it rigidly} rotating cloud
or a turbulent pre-stellar core. 

\textcolor{black}{
The study of a maser outflow in the high-mass star forming region W75N 
equally conjectures that "short-lived outflows in massive protostars 
are probably related to episodic increases in the accretion rates, as observed 
in low-mass star formation"~\citep{Carrasco_sci_348_2015}. 
}
Additionally, a luminosity outburst of the massive ($\approx 20\, \rm M_\odot$) young star S255IR-NIRS3 
was reported in~\citet{fujisawa_atel_2015} by means of 6.7 GHz methanol maser emission.
This emission line has been discovered 
by~\citet{menten_apj_380_1991} and constitutes today a well known-tracer of 
high-mass star forming regions~\citep[see][and references therein]{bartkiewicz_aa_587_2016}.
\textcolor{black}{Recent observation of the same object show brightness variations that resemble strongly}
FU-Orionis-type outbursts~\citep{stecklum_ATel_2016}. 

Motivated by the above listed numerical studies and observational arguments, we continue 
to investigate the burst phenomenon in the high-mass regime of star 
formation. We follow existing models 
showing accretion spikes in high-mass star formation, further analyze their nature 
in the context of the star-disc evolution and conjecture on possible observational implications.  
This study is organized as follows. In Section~\ref{sect:method}, we 
review the methods that we utilise to carry out our high-resolution 
\textcolor{black}{self-gravity radiation-hydrodynamical} simulation of the formation and evolution of a disc surrounding a 
growing present-day massive protostar generated by the collapse of a {\it 
non-rigidly} rotating pre-stellar core. Our outcomes are 
presented and discussed in Section~\ref{sect:results}. Particularly, our model 
also generates such outbursts and we show that they are caused by the rapid 
migration of disc fragments onto the protostar. Finally, we conclude on their 
significances in Section~\ref{section:cc}.


\section{Numerical simulation}
\label{sect:method}

\begin{figure*}
        \centering
        \begin{minipage}[b]{ 0.44\textwidth}
                \includegraphics[width=1.0\textwidth]{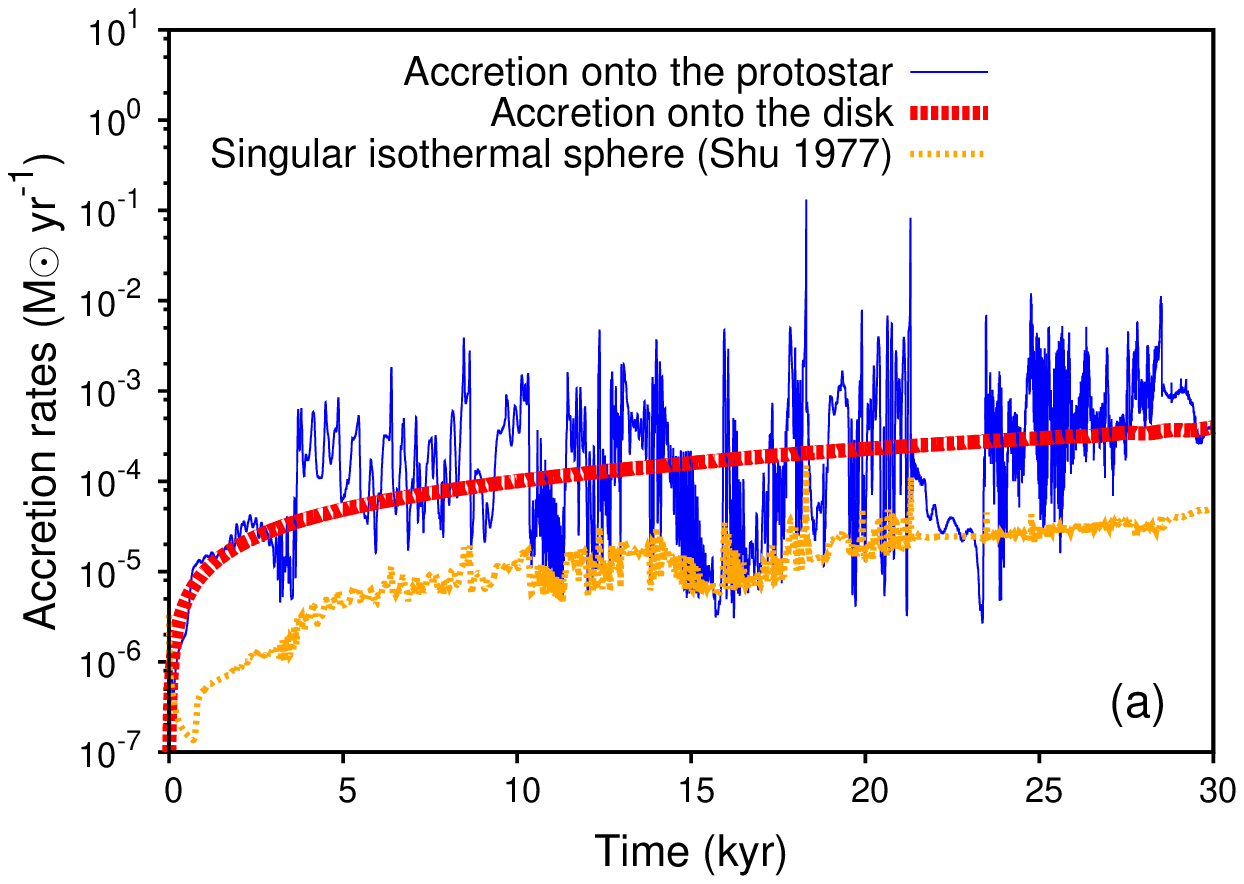}
        \end{minipage}  
        \begin{minipage}[b]{ 0.44\textwidth}
                \includegraphics[width=1.0\textwidth]{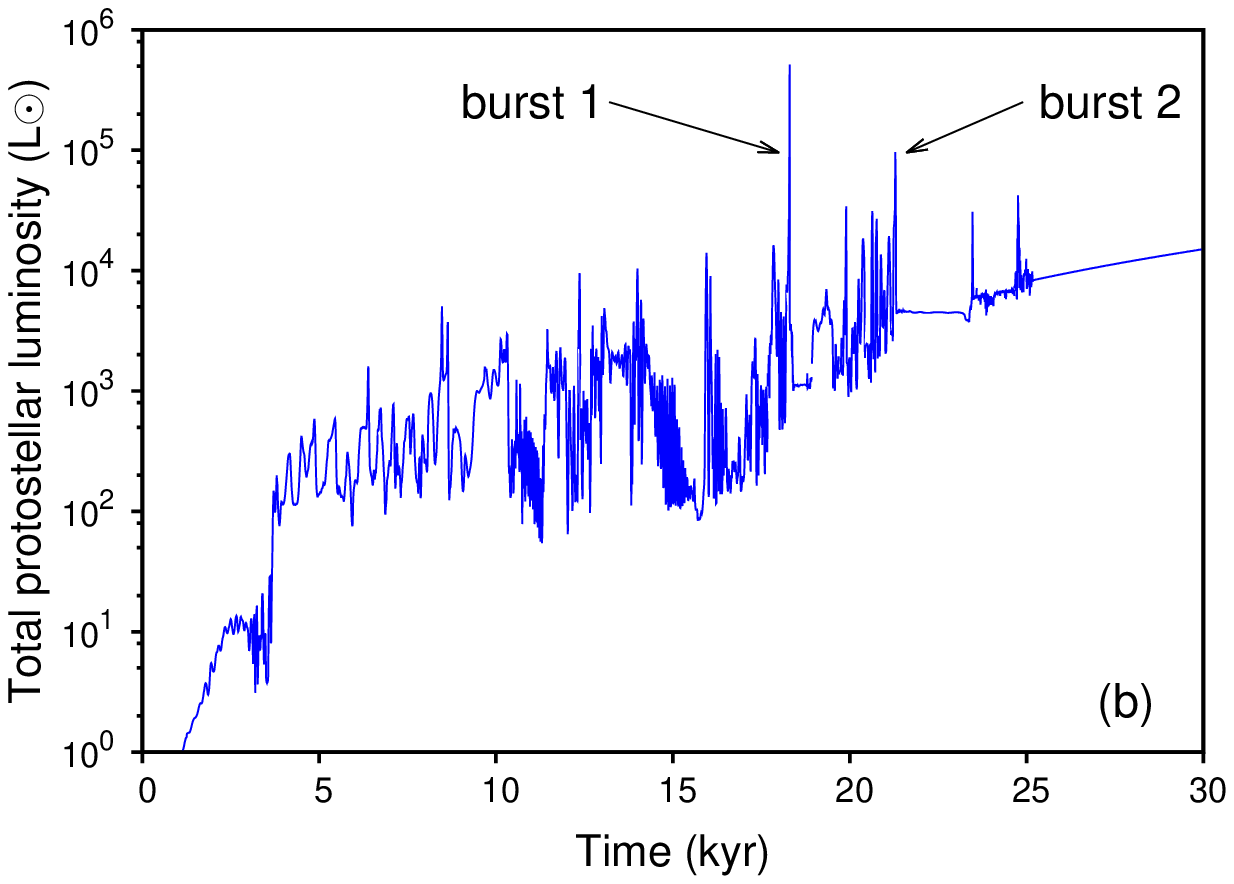}
        \end{minipage}                  
        \caption{ 
                 Left: accretion rate onto the protostar 
                 and mass infall rate onto the disc (in $\rm M_{\odot}\, \rm yr^{-1}$).   
                 Right: total luminosity of the protostar (in $\rm L_{\odot}$).  
                 }      
        \label{fig:gen_plots}  
\end{figure*}

We perform a 3D numerical radiation hydrodynamics simulation with the {\sc 
pluto} code~\citep{mignone_apj_170_2007,migmone_apjs_198_2012} that has been 
augmented with several physics modules for (i) self-gravity of the gas and (ii) 
a careful treatment of the proto-stellar irradiation 
feedback~\textcolor{black}{\citep{kuiper_aa_511_2010,kuiper_aa_555_2013}}. Our model has been 
carried out using a spherical coordinate system (r,$\theta$,$\phi$)  mapped 
with a grid of size $[\textcolor{black}{r_{\rm min}},r_{\rm max}]\times[0,\pi/2]\times[0,2\pi]$ that 
is made of $128\times21\times128$ cells, respectively. The \textcolor{black}{grid resolution} expands 
logarithmically in the radial direction, scales as $\cos(\theta)$ in the polar 
direction, is uniform in the azimuthal direction and assumes mid-plane symmetry 
about $\theta=\pi/2$, such that it allows us to reach sub-AU spatial 
resolution in the inner region of the disc. 
The dynamics of the collapse and the accretion flow is calculated in the frame 
of reference of the fixed protostar that is considered to be evolving inside a 
\textcolor{black}{static, semi-permeable sink-cell of initial mass 
$\approx10^{-3}\, \rm M_{\odot}$ with a radius $r_{\rm min}=10\,\rm AU$ 
such that the mass flux through it represents the protostellar accretion rate $\dot{M}$ 
onto the stellar surface. The creation of other sink cells is not allowed. } As 
described in~\citet{kuiper_apj_722_2010}, we utilise $\dot{M}$ to self-consistently update 
the stellar mass and interpolate the pre-calculated proto-stellar evolutionary 
tracks of~\citet{hosokawa_apj_691_2009} to use them as boundary conditions for 
the radiation transport module.

We follow the gravitational collapse of a $M_{\rm c}=100\, \rm M_{\odot}$ 
pre-stellar core of  outer radius $r_{\rm max}=0.1\, \rm pc$. 
The radiation is regulated using the dust opacity prescription 
of~\citet{laor_apj_402_1993} while the constant gas opacity is taken to be 
$\kappa_{\rm g}=0.01\, \rm g\, \rm cm^{-2}$. This hybrid radiation transport 
method allows us to carefully treat disc thermodynamics and to generate the dust sublimation 
front in massive protostellar accretion discs~\citep[cf.][]{vaidya_apj_702_2009}. 
\textcolor{black}{
Our pre-stellar core has the canonical radial density distribution $\rho(r)\propto 
r^{-3/2}$. We do not consider a rigidly-rotating core but impose a radial angular momentum 
distribution $\Omega(R) \propto R^{-3/4}$ with $R = r \sin(\theta)$ the 
cylindrical radius. 
We impose a ratio of rotational to gravitational energy $\beta=4\%$,  
its initial temperature is uniformly taken to $T_{\rm g}=10\, \rm K$ and we set its pressure assuming 
that the gas is at solar metallicity and obeys an ideal equation of state.} The dust 
temperature $T_{\rm d}$ is considered as equal to the gas temperature 
$T_{\rm g}=T_{\rm d}$. The collapse of the core and the evolution of the 
accretion disc is followed during $30\, \rm kyr$.


\section{Results}
\label{sect:results}


\subsection{Episodic accretion of dense gaseous clumps}
\label{sect:fall}

The midplane density field from \textcolor{black}{the} simulation (in $\rm g\, \rm cm^{-3}$) is 
shown in Fig.~\ref{fig:disk_density_plots} at different evolutionary times (in 
$\rm kyr$). While panel (a) represents an overview of the whole disc \textcolor{black}{region (where material orbits the star with
approximately Keplerian velocity)}, panels 
(b)-(d) are zooms focusing on the inner region of the disc. The first snapshot 
(a) shows the circumstellar medium of the protostar at a time $18.10\, \rm kyr$. 
In addition to the gas that is being 
accreted in the innermost $50\, \rm AU$ of the disc, it develops inhomogeneities 
by gravitational instability that take the form of spiral arms terminated by 
dense blobs of gas orbiting at radii $450$$-$$800\, \rm AU$ from the 
protostar. One of those Truelove-resolved arms~\citep[$\lambda/\Delta\sim 6-8$, where 
$\lambda$ is the Jeans length and $\Delta$ is the grid cell size, see][]{truelove_apj_495_1998} 
experiences a local increase in density and 
develops an overdense clump of circumstellar material at a radial distance of 
$\approx 200\, \rm AU$ to the central star. This is how fragmentation occurs at those 
radii ($\ge 150\, \rm AU$) predicted by the 
analytic study of~\citet{kratter_mnras_373_2006}, despite the 
fact that only marginal fragmentation for discs like the one around our 
$\approx 4.5\, \rm M_{\odot}$ protostar is foreseen (see also Fig.~\ref{fig:zoom_burst}b).

By means of angular momentum transfer to the closest spiral arm, the clump migrates 
towards the protostar. 
\textcolor{black}{
At the onset of migration (Fig.~\ref{fig:disk_density_plots}a), the closest \textcolor{black}{distance} to which 
the clump can fall, assuming angular momentum conservation, is $\sim2000\, \rm R_{\odot}$, 
and it decreases to $\sim 260\, \rm R_{\odot}$ (Fig.~\ref{fig:disk_density_plots}b) before to 
become comparable to the stellar radius of $\sim90\, \rm R_{\odot}$ shortly before entering the sink cell. }
This value is certainly an upper limit because 
the angular momentum of the clump decreases during its inward migration thanks to the 
gravitational interaction with the closest spiral arm (the gravitational torque acting on 
the clump is negative), as was shown in application to low-mass star formation 
by~\citet{voroboyov_apj_650_2006}. 
\textcolor{black}{
As a consequence, we expect the clump to fall directly onto the star.
However, if the gas temperature in the clump interior exceeds 2000~K, the clump will
further contract owing to dissociation of molecular hydrogen. In this case, not captured
by our numerical simulations due to limited numerical resolution, 
its subsequent evolution will depend on the amount of rotational energy in the clump.
By extrapolation of the clump properties
formed in discs around solar-mass stars showing that about half of the clumps 
are rotationally supported \citep[see][]{vorobyov_2016}, we expect that part of the 
clump material will retain in the form of a disc/envelope around the newly formed protostar. 
This material can still be lost onto the central protostar via the Roche lobe 
creating an accretion burst though of lesser amplitude, 
as was shown in \citet{nayakshin_mnras_426_2012}. 
Furthermore, high-resolution studies of low-mass and primordial star formation showed that 
further fragmentation may happen within in the innermost disk regions occupied by our sink cell, 
likely triggered by already existing outer 
fragments or efficient $H_2$ cooling and collisionally induced emission~\citep[see][]{meru_mnras_454_2015,greif_mnras_424_2012}.
If fragmentation in the inner disk occurs also in the case of high-mass star formation, this may provoke the formation of a pattern of very small clumps which complex mutual 
gravitational interactions can make them merging together, orbiting onto the protostar 
or being dynamically ejected away. This would modify the physics of accretion and 
the accretion luminosity of the protostar, however, heavy clumps migrating very 
rapidly will still fall directly onto the stellar surface. 
} 
To follow this process in detail, however, a much smaller (than 10~AU) sink cell and, as a consequence, much 
longer integration times are needed. We leave this investigation for a subsequent \textcolor{black}{study}.

At time $18.28\, \rm kyr$, the clump is at about $60\, \rm AU$ from the 
protostar (Fig.~\ref{fig:disk_density_plots}b) and its density reaches 
$\ge 10^{-9}\, \rm g\, \rm cm^{-3}$. \textcolor{black}{The self-gravitating 
clump adopts a slightly elongated shape due to the gravitational influence 
of the massive protostar}. At time $18.30\, \rm kyr$ 
(Fig.~\ref{fig:disk_density_plots}c), the stretched clump starts wrapping around 
the sink cell. A part of its material is accreted onto the protostar, and, 
consequently, it affects the star's accretion luminosity 
$L_{\rm acc}\propto G M_{\star} \dot{M}/R_\star$ (see Section~\ref{sect:burst}). 
When the clump starts migrating towards the star its radius is 20~AU. As the clump approaches 
the sink cell, its mass has decreased to $\simeq0.7~\rm M_\odot$ and its Hill radius decreases
from approximately 35~AU to $\la 10$~AU, causing the clump to lose part of its mass
in the diffuse outer envelope. Finally, about $0.55~\rm M_\odot$ is accreted through the sink cell. 
After time $18.32\, \rm kyr$ 
(Fig.~\ref{fig:disk_density_plots}d), the structure of the density field around 
the sink cell is that of a spiral of unaccreted dense gas material. 

This scenario is episodic in the disc evolution and repeats itself at a time
about $21.00\, \rm kyr$ when another clump is similarly accreted. The time
interval between recurrent accretion bursts can be determined as 
the time needed to replenish the disc material lost during the last accretion event 
$t_{\rm repl} = M_{\rm cl}/\dot{M}_{\rm
infall}$, where $M_{\rm cl}$ is the clump mass and $\dot{M}_{\rm infall}$ the mass
infall rate onto the disc from the collapsing envelope (see Section 3.2). For $M_{\rm cl}\sim
0.55~\rm M_\odot$ and $\dot{M}_{\rm infall}=(2-3)\times 10^{-4}~\rm M_\odot$~yr$^{-1}$,
typical values for the time of the bursts, the characteristic time of recurrent outbursts 
is $t_{\rm repl}\approx2-2.5$~kyr. If $t_{\rm repl}$ is regarded as the
time required to increase the disc mass to the limit when the disc fragments, i.e.,
as the fragmentation timescale,
then the clump migration timescale is much shorter than
the fragmentation timescale $t_{\rm repl}$. This is indeed evident from
Figure~\ref{fig:disk_density_plots}.

%
%
%
%

\subsection{Accretion-induced strong outbursts}
\label{sect:burst}

The accretion rate $\dot{M}$ onto the protostar exhibits rapid variations after 
the disc formation at about $4\, \rm kyr$ (see Fig.~\ref{fig:gen_plots}a). 
Comparing $\dot{M}$ measured at $\textcolor{black}{r_{\rm min}}=10$~AU with the smooth infall 
rate $\dot{M}_{\rm infall}$ from the envelope onto the disc measured at 
$r=3000\, \rm AU$, it becomes evident that the oscillations of $\dot{M}$ from 
about $10^{-6}$ to about $10^{-3}\, \rm M_{\odot}\, \rm yr^{-1}$ are caused by 
the disc gravitational instability when filaments and spiral arms of different 
density and length converge towards the protostar. This correlation between the 
strength of disc gravitational instability and the protostellar accretion 
variability is well known in low-mass star 
formation~\citep{dunham_apj_747_2012,vorobyov_apj_805_2015}. The envelope, on 
the other hand, has a smooth density and velocity structure, and its infall rate 
that gradually increases with time \textcolor{black}{for 
the initial free-fall time of the mass reservoir.}. Note also that the self-similar 
solution for the a singular isothermal sphere~\citep{shu_apj_214_1977} 
computed as the mean isothermal infall rate at $r=3000\, \rm AU$, underestimates the accretion onto 
the protostar by about an order of magnitude (Fig.~\ref{fig:gen_plots}a), as previously 
noticed in the context of low-mass and high-mass star formation, 
see~\citet{vorobyov_apj_704_2009} and~\citet{benerjee_apj_660_2007}, respectively.

\begin{figure}
        \centering              
        \begin{minipage}[b]{ 0.435\textwidth}
                        \includegraphics[width=0.98\textwidth]{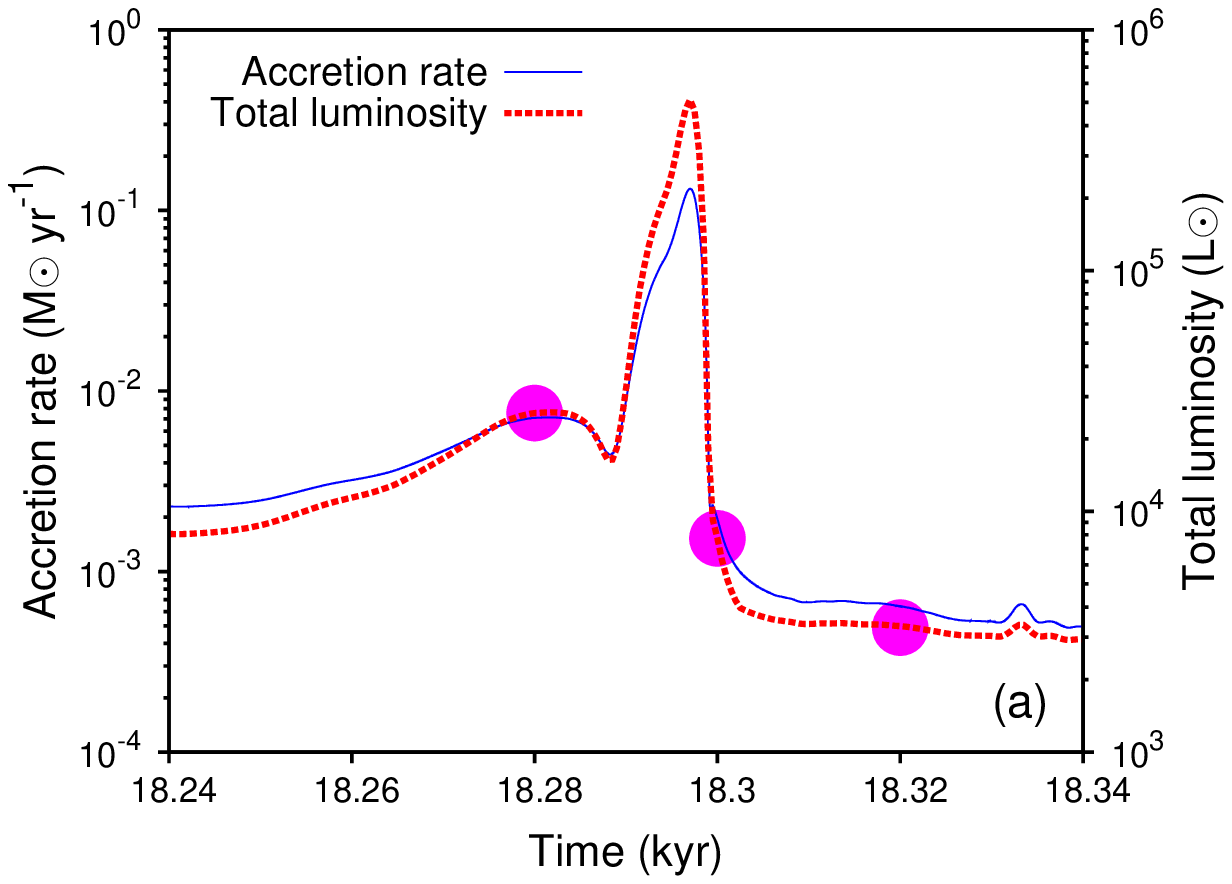}
        \end{minipage}  
        \begin{minipage}[b]{ 0.45\textwidth}
                        \includegraphics[width=0.98\textwidth]{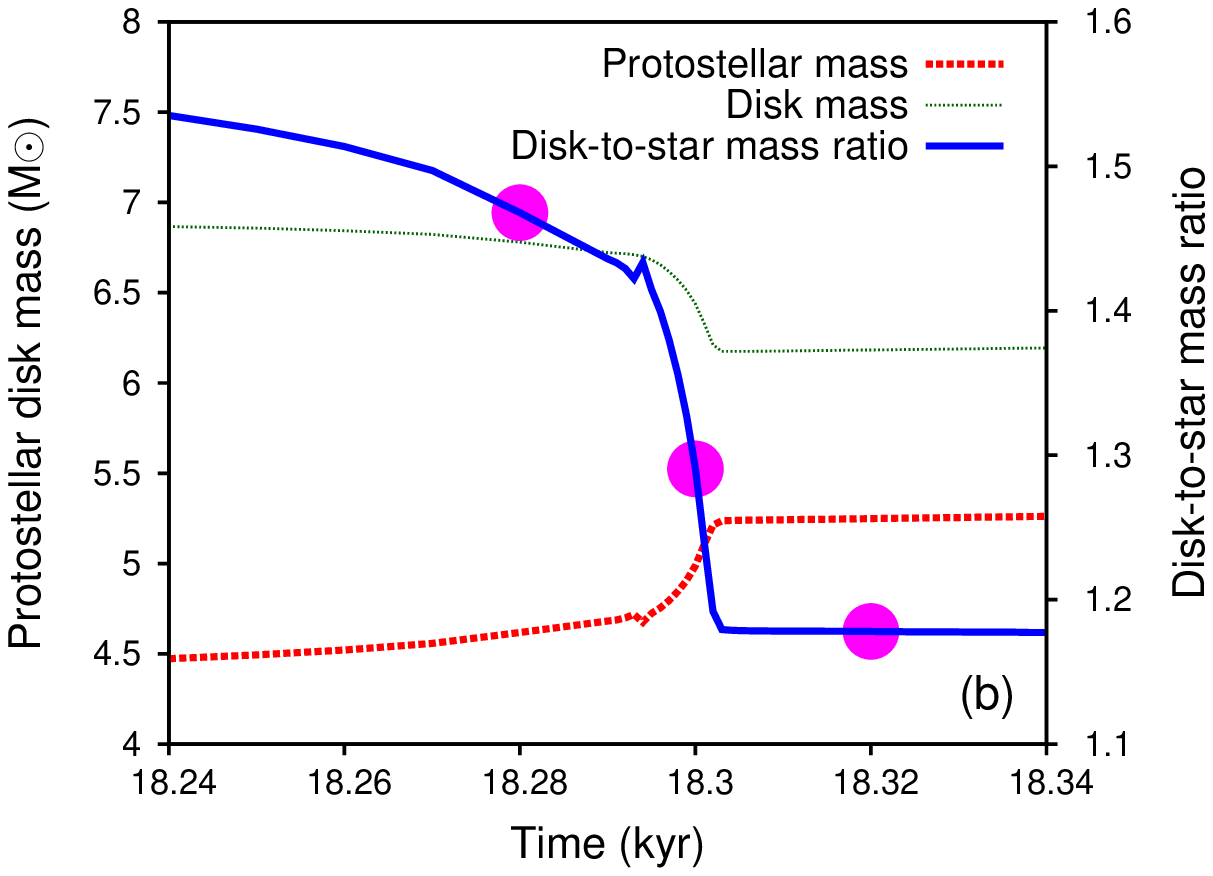}
        \end{minipage}        
        \caption{ 
                 Upper panel: As in Fig.~\ref{fig:gen_plots} during the time 
interval corresponding to the first accretion-induced outburst. Bottom panel: 
Protostellar mass ($\rm M_{\odot}$) and disc-to-star mass ratio, for a time 
interval excluding the initial gravitational collapse where no disc has formed 
yet. The magenta dots mark the times of the zooms in Fig.~\ref{fig:disk_density_plots}b-d.      
                 }      
        \label{fig:zoom_burst}  
\end{figure}

At times $18.29$~kyr and $21.30\, \rm kyr$ the accretion of parts of the heaviest 
clumps occurs. The total luminosity of the protostar, i.e., the sum of its photospheric 
luminosity $L_{\star}$ and its accretion luminosity 
$L_{\rm acc}$ rises simultaneously (see Fig.~\ref{fig:gen_plots}b).
The first luminosity outburst is particularly strong reaching in magnitude 
$5\times 10^5~L_\odot$. The second burst is notably weaker because the 
\textcolor{black}{intrinsic luminosity of the protostar has greatly 
increased after the first one.} 
These accretion and luminosity outbursts are a direct consequence of the accretion 
of in-spiraling clumps that fall onto the protostar because of the loss of angular 
momentum that has been exchanged with other structures of the disc such as 
spiral arms, filaments, and arcs. 
In Fig.~\ref{fig:zoom_burst} we present various disc and stellar properties around the time of the first 
burst.  The accretion rate exhibits a rapid increase 
from $\dot{M} \sim 10^{-3}~M_\odot$~yr$^{-1}$ to $\dot{M} \approx 10^{-1}\, \rm M_{\odot}\, \rm yr^{-1}$ 
over a time interval of $\approx10\, \rm yr$, while the protostar experiences a flare reaching 
$\ge 5 \times 10^{5}\, \rm L_{\odot}$ (Fig.~\ref{fig:zoom_burst}a). As the 
protostar accretes the clump, its mass increases while the disc mass decreases 
by the amount of mass of the accreted clump.  
This results in a remarkable step-like decrease of the \textcolor{black}{disc-to-star mass} ratio 
(see Fig.~\ref{fig:zoom_burst}b), making the disc temporarily stable to 
gravitational fragmentation. It takes a \textcolor{black}{few} several kyr for the disc to accrete 
additional mass from the collapsing envelope and produce another similar fragmentation episode. 
This phenomenon repeats for as long as there is enough material in the envelope to replenish the disc
mass, as was previously found for low-mass~\citep{vorobyov_apj_723_2010} and 
primordial star formation~\citep{vorobyov_apj_768_2013, hosokawa_2015}. 

\section{Discussion and conclusion }
\label{section:cc}

We have confirmed the presence of strong peaks in the variable 
accretion history of high-mass protostars, already present in the literature 
\textcolor{black}{for a wide range of initial conditions of the pre-stellar core. 
All models have a standard initial density distribution $\propto r^{-3/2}$ 
that is assumed in the absence of corresponding observations. All rotating models 
preceding our study assumed cores in sold-body rotation. Turbulent 
radiation-hydrodynamics simulations include~\citep{krumholz_apj_656_2007,peters_apj_711_2010,seifried_mnras_417_2011}, 
other studies neglected turbulence. Initial conditions of $M_{\rm c}$ and $\beta$-ratio 
are $100$$-$$200\, \rm M_{\odot}$ with $2\%$~\citep{krumholz_apj_656_2007}, 
$100\, \rm M_{\odot}$ with $2\%$~\citep{krumholz_sci_323_2009,kuiper_apj_732_2011},  
$1000\, \rm M_{\odot}$ with $5\%$~\citep{peters_apj_711_2010}, 
$100\, \rm M_{\odot}$ with $4$$-$$20\%$~\citep{peters_apj_711_2010} and 
$100;200\, \rm M_{\odot}$ with $10.5;5.3\%$~\citep{klassen_arXiv160307345K}, 
respectively, while we use $100\, \rm M_{\odot}$ with $4\%$. 
} 
Our higher spatial resolution in the inner disc allows us to explicitly 
capture the fall of forming circumstellar clumps rather than modelling overdense 
filaments in the disc that wrap onto the star. 
It enables to study this process, well-known in the low-mass and primordial regime 
of star formation as the so-called episodic accretion-driven outbursts, which we 
conclude to be also present in the high-mass regime of contemporary star formation. 
\textcolor{black}{
To confirm this result, we have performed preliminary simulations varying the 
initial rotation curves of the pre-stellar core.
}
In the subsequent studies, numerical simulations with a smaller sink cell 
are needed to determine more in detail the final fate of the accreted clumps. However, we expect that the 
protostellar accretion history remains quantitatively similar in models with
sink radii varied by a factor of two, as was earlier shown for low-mass star 
formation~\citep[see][and references therein]{vorobyov_apj_805_2015}.

\textcolor{black}{We obtain} the protostellar accretion history 
which is highly time-variable and shows sudden accretion spikes, as was 
found in several previous models of massive-star-forming pre-stellar 
cores 
and interpreted as caused by azimuthal 
asymmetries in the accretion flow. The higher spatial resolution of 
our model reveals that, in addition to variable accretion indeed caused by the asymmetric 
character of its disc, the growing 
massive protostar can accrete material of gaseous clumps formed in spiral arms owing
to disc gravitational fragmentation, 
which rapidly migrate towards the protostar and induce luminous protostellar outbursts.  
A similar rapid migration of dense clumps is present in 
gravitationally unstable discs around solar-mass 
stars, in primordial discs around the first stars and when high-mass 
planets form~\citep{2011MNRAS.416.1971B,vorobyov_aa_552_2013}.  
Our work indicates that this also applies to the 
high-mass regime of star formation, which implies a change in the paradigm which, to the best of our knowledge, considered
episodic accretion-induced protostellar outbursts as inherent to low-mass and 
primordial star formation only. Our study shows that it concerns star 
formation in general, for a wide range of the physical properties such as the initial mass, 
$\beta$-ratio, angular momentum distribution or chemical composition of the 
parent environment in which stars form. 

It supports the consideration of star formation as a process ruled by a common set 
of mechanisms leading to circum-protostellar structures similarly 
organized, but scaled-up with respect to each other as a function of the initial 
properties of their parental pre-stellar cores, as observationally suggested~\citep{shepherd_apj_472_1996,
fuente_aa_366_2001,testi_2003,keto_mnras_406_2010,johnston_apj_813_2015}. 
Finally, we propose to consider flares from high-mass protostellar objects as a 
possible tracer of the fragmentation of their accretion discs. This may apply to 
the young star S255IR-NIRS3 that has recently been associated to a 6.7 GHz 
methanol maser outburst~\citep{fujisawa_atel_2015,stecklum_ATel_2016}, but also to the 
other regions of high-mass star formation from which originated similar 
flares~\citep{menten_apj_380_1991} and which are showing evidences of 
accretion flow associated to massive protostars, see e.g. 
\textcolor{black}{in} W3(OH)~\citep{hirsch_apj_757_2012}, 
W51~\citep{keto_apj_678_2008,zapata_apj_698_2009} and 
W75~\citep{Carrasco_sci_348_2015}.



\section*{Acknowledgements}

\textcolor{black}{ We thank the anonymous reviewer for his valuable comments which 
improved the quality of the paper. }
This study was conducted within the Emmy Noether research group on "Accretion Flows 
and Feedback in Realistic Models of Massive Star Formation" funded by the German Research 
Foundation under grant no. KU 2849/3-1. E.I.V. acknowledges support from the 
Austrian Science Fund (FWF) under research grant I2549-N27 and  RFBR grant 14-02-00719.


\bibliographystyle{mn2e}

\footnotesize{
\bibliography{grid}
}


\end{document}